\documentclass[10pt]{article}

\usepackage{amsmath}
\usepackage{amssymb}

\usepackage{graphicx}

\usepackage{cite}

\usepackage{color} 

\usepackage[labelfont=bf,labelsep=period,justification=raggedright]{caption}
\usepackage{subcaption}
\usepackage{float}
\usepackage{multirow}
\makeatletter
\renewcommand{\@biblabel}[1]{\quad#1.}
\makeatother

\date{}

\begin{document}

\begin{flushleft}
{\Large
\textbf{Table-Top Molecular Communication: Text Messages Through Chemical Signals}
}
\\
Nariman Farsad$^{\ast \ddagger}$, 
Weisi Guo$^\dagger$,
Andrew W. Eckford$^\ast$ 
\\
$^*$ Dept. of Electrical Engineering and Computer Science, York University, Toronto, Ontario, Canada
\\
$^\dagger$ School of Engineering, University of Warwick, Coventry, UK
\\
E-mail: nariman@cse.yorku.ca (NF); weisi.guo@warwick.ac.uk (WG); aeckford@yorku.ca (AWE)
\\
$\ddagger$ Corresponding author
\end{flushleft}

\section*{Abstract}
In this work, we describe the first modular, and programmable platform capable of transmitting a text message using chemical signalling -- a method also known as molecular communication. This form of communication is attractive for applications where conventional wireless systems perform poorly, from nanotechnology to urban health monitoring. Using examples, we demonstrate the use of our platform as a testbed for molecular communication, and illustrate the features of these communication systems using experiments. By providing a simple and inexpensive means of performing experiments, our system fills an important gap in the molecular communication literature, where much current work is done in simulation with simplified system models. A key finding in this paper is that these systems are often nonlinear in practice, whereas current simulations and analysis often assume that the system is linear. However, as we show in this work, despite the nonlinearity, reliable communication is still possible. Furthermore, this work motivates future studies on more realistic modelling, analysis, and design of theoretical models and algorithms for these systems.


\section*{Introduction}
The need to convey information over a distance has always been an important part of human society. Many techniques have been used throughout history, such as semaphores, fire beacons, smoke signals, carrier birds, electrical signals, and electromagnetic waves. Although modern telecommunication systems rely almost entirely on electrical or electromagnetic signals, there are still many applications were these technologies are not convenient or appropriate. For example, use of electromagnetic wireless communication inside networks of tunnels, pipelines, or unpredictable underwater environments, can be very inefficient. As another example, at extremely small dimensions, such as between micro- or nano-scaled devices \cite{aky2008}, electromagnetic communication is challenging because of constraints such as the ratio of the antenna size to the wavelength of the electromagnetic signal. 

Inspired by nature, one possible solution to these problems is to use {\em chemical signals} as carriers of information, which is called {\em molecular communication}. For example, chemical signals are used for inter-cellular and intra-cellular communication at micro- and nano-scales \cite{alb07}, and pheromones are used for long range communication between members of the same species such as social insects \cite{ago92}. Therefore, chemical signals can be used for communication at both macroscopic and microscopic scales. Moreover, molecular communication signals are biocompatible, and require very little energy to generate and propagate. These properties makes chemical signals ideal for many niche applications, where the use of electromagnetic signals are not possible or not desirable.      

At microscopic scales, chemical signalling has been proposed as an effective solution for communication between engineered micro- or nano-scaled devices \cite{hiy05, eck07, hiy10NanoCom, nak12} such as lab-on-a-chip devices \cite{far12NanoBio} and body area sensor networks \cite{ata12}. At macroscopic scales, use of very primitive molecular communication has been proposed in robotics for distress signalling by defective robots \cite{pur05}, estimating the size of a swarm of robots (quorum sensing) \cite{pur10}, and as chemical trails for robot guidance \cite{sou08, sou08b}. 

Despite all this recent work, there have been few practical demonstrations of molecular communication systems that can be used to transfer data and messages, at either the macroscopic or microscopic scales. For example, at microscopic scales, one of the major obstacles in implementing molecular communication is the tedious, laborious and expensive nature of wet lab experimentation. As a result, a large body of work on the theoretical aspects of microscopic molecular communication systems has been developed \cite{moo09b, eck10, nak10, mah10, pie10, far11NanoCom, gun11, sri12, far12NANO, kim12, pie13}, without any physical implementation of a fully functional communication device. Similarly at macroscopic scales, the potential for data communication has not been explored: the systems we cited above use chemical markers for localization or navigation only.

In this work, we implement a macroscopic molecular communication system for transmitting a brief text message using chemical signals. To our knowledge, this is the first implementation of a macroscopic data communication system using chemical signals, and it is one of very few implementations of molecular communication at any dimension. Our system is an ideal experimental platform for interdisciplinary researchers to gain experience in the growing field of molecular communication: 
it is relatively inexpensive (in the hundreds of US dollars), and compact (fitting literally on a ``table top''), requiring no supporting hardware or surrounding laboratory infrastructure. 
Moreover, 
our system provides a demonstration platform for molecular communication at macroscopic scales:
in this form, our system could be used to evaluate data transfer between robots, and in environments that electromagnetic communication is not possible or desirable. Whilst the test-bed itself is a macroscopic scale demonstration of molecular communication, the off-the-shelf equipment can be miniaturized easily in future generation platforms until molecular communication at microscopic scales is achieved.   

In designing this simple and robust experimental apparatus, our vision is to provide a bridge from the rapidly growing body of modelling and theoretical work in molecular communication, to the practical applications that will demonstrate the transformative power of the concept: 
microscopically, in medical diagnostics and targeted drug delivery; and macroscopically, in sewer systems, pipelines, smart cities, and disaster search and rescue operations. This is our first contribution.

As the second contribution, we study the effects of flow on transporting chemical signals within our platform. We generate different types of flow using bladed and bladeless fans at different speeds. We then analyse and report the effects of different types of flow on overall impulse response of the system.

\section*{Materials and Methods}
In this work we implement a simple, robust, cost-effective communication system that uses chemical signals for carrying information from a transmitter to a receiver. To test this system, we use it to send a short text message: this is a familiar application, as billions of Short Message Service (SMS) text messages are sent daily by mobile users \cite{ogr12}.
To design and develop our system, we use the following criteria:
\begin{itemize}
\item The end product must be inexpensive to build. This would make the platform readily available for many different research and development projects with limited amount of founding.
\item The designed system must be simple and robust, much the same as the telegraph, the ancestor of modern telecommunication systems. While we are proposing a first-generation device, the simple and robust design would help in the adoption of the platform in different applications.
\item The developed system must be easily modifiable and programmable. Again this is an important criterion for future expansions and adoption to different applications. 
\end{itemize}

\begin{figure}[!ht]
	\begin{center}
		\includegraphics[width=5in]{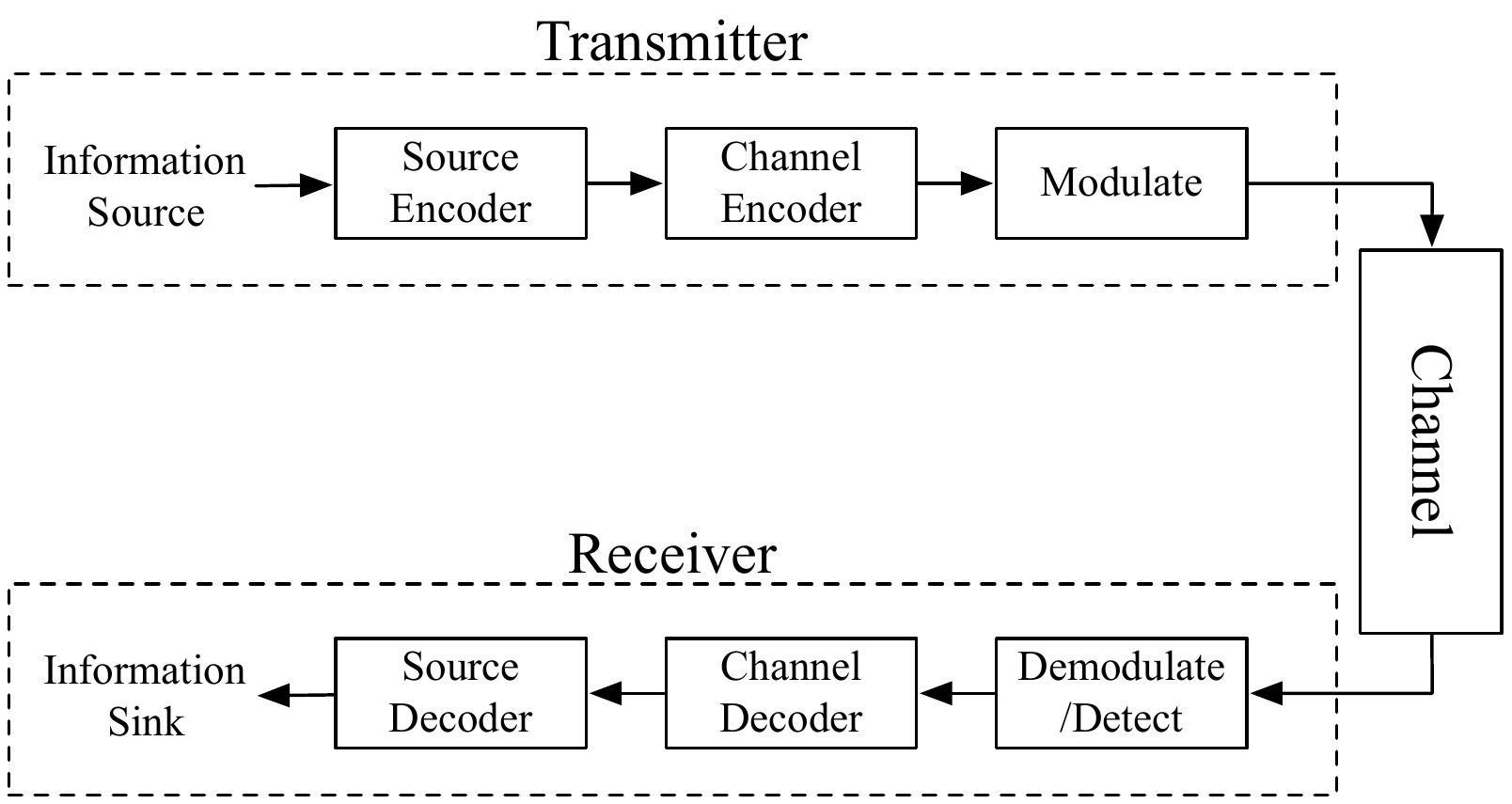}
	\end{center}
	\caption{\label{fig:commDiag} {\bf Block diagram of a typical communication system.}}
\end{figure}

Any communication system can be broken down into three major parts: the transmitter, the receiver, and the channel. Figure \ref{fig:commDiag} shows the block diagram representation of these three modules and their submodules. The transmitter has some information that it wants to transmit to a receiver. Any discrete message can be represented with a string of binary numbers, so the transmitter uses a source encoder to encode the information message as a binary sequence. The transmitter can also add error-correcting codes using the channel encoder block, which essentially introduces redundancy by adding extra bits. The receiver can use the added redundancy to mitigate the errors that may be introduced by the channel. Finally, the transmitter must modulate the channel symbols (i.e. the output of the channel encoder) onto a carrier signal and release the signal for propagation in the channel.

The channel is the environment in which the transmitted signal propagates from the sender to the receiver. For example, a channel could be a wire where electrical signals propagate or air where electromagnetic signals propagate. The channel may introduce noise into the system, where the noise is any distortion that results in degradation of the signal at the receiver. For example, the noise can result from the signal fading as the it propagates, or interference from other signals. Noise can also be introduced by the transmitter and the receiver themselves (e.g. thermal noise in the electronic components). When the transmitted signal arrives at the receiver, the receiver must first demodulate and detect the channel symbols. The estimated channel symbols are then decoded using a channel decoder, where some of the errors introduced by the transmitter, the channel and the receiver may be corrected. The output of the channel decoder goes through source decoder, where the receiver estimates what information the transmitter has sent. If this estimation is correct, then the communication session has been successful.

Based on these criteria, we present our design for the transmitter, the receiver, and channel propagation in the next couple of subsections.

\subsection*{Transmitter Design}
The transmitter takes an input text message from a user. It then converts the text message into a sequence of binary bits and modulate them on a chemical signal for propagation in the channel. To control all transmission operations, we use the Arduino Uno open-source electronics prototyping platform, which is an ATmega328 based microcontroller board. For text entry, we use the 16x2 character LCD Shield Kit from Adafruit. The LCD is an add-on module for the Arduino microcontroller board, which also has six push buttons. We wrote a program for the Arduino microcontroller which employs the LCD and its buttons for text entry by the user.

To convert the text message to a binary sequence, we use the International Telegraph Alphabet No. 2 (ITA2) standard \cite{blueBook}, where every letter is represented using five bits. For example, the letter ``E'' is represented by a five bit sequence ``10000''. For simplicity, in this work we do not use any error-correcting code. Therefore, the five bit encoded letters are passed to the modulator block of the transmitter for modulation and transmission to the channel.
\begin{figure}[!h]
	\begin{center}
		\includegraphics[width=3.2in]{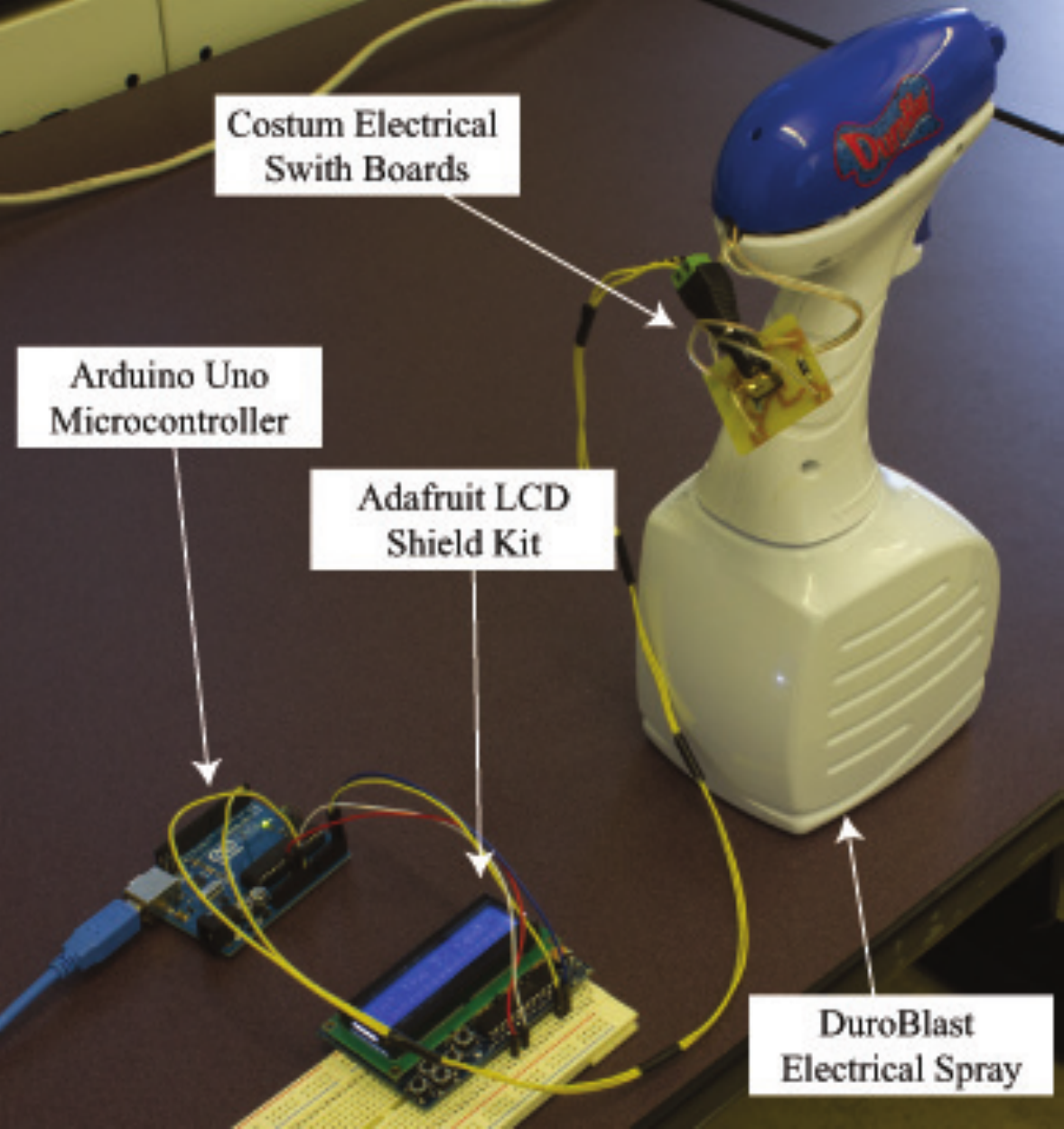}
	\end{center}
	\caption{\label{fig:TransPic} {\bf The transmitter components.}}
\end{figure}

To modulate the channel symbols into chemical signals, we use an electronic spray called DuroBlast made by Durotech Industries. The DuroBlast electronic spray has a battery operated electrical pump that can spray a wide variety of liquid chemicals that can be stored inside its container. We designed a custom electrical switch board that can be used to control the spray from the Arduino microcontroller board. By programming the Arduino microcontroller board, any type of modulation can be implemented through controlled set of sprays. Figure \ref{fig:TransPic} shows our transmitter setup with all of its subcomponents.

\subsection*{Receiver Design}
To design the receiver, a sensor is required that is capable of detecting a chemical signal. The data from the sensor is processed by the demodulation and detection algorithms, and finally decoded into text. Again we use the  Arduino Uno open-source microcontroller for programming and controlling all the receiver operations. The Arduino Uno board has a 10-bit analog to digital converter that can be used to read the sensor data. The demodulation and detection block and the source decoder block can then be programmed into the microcontroller, and the resulting detected text message can be displayed on a computer screen using serial port.
\begin{figure}[!h]
	\begin{center}
		\includegraphics[width=3.2in]{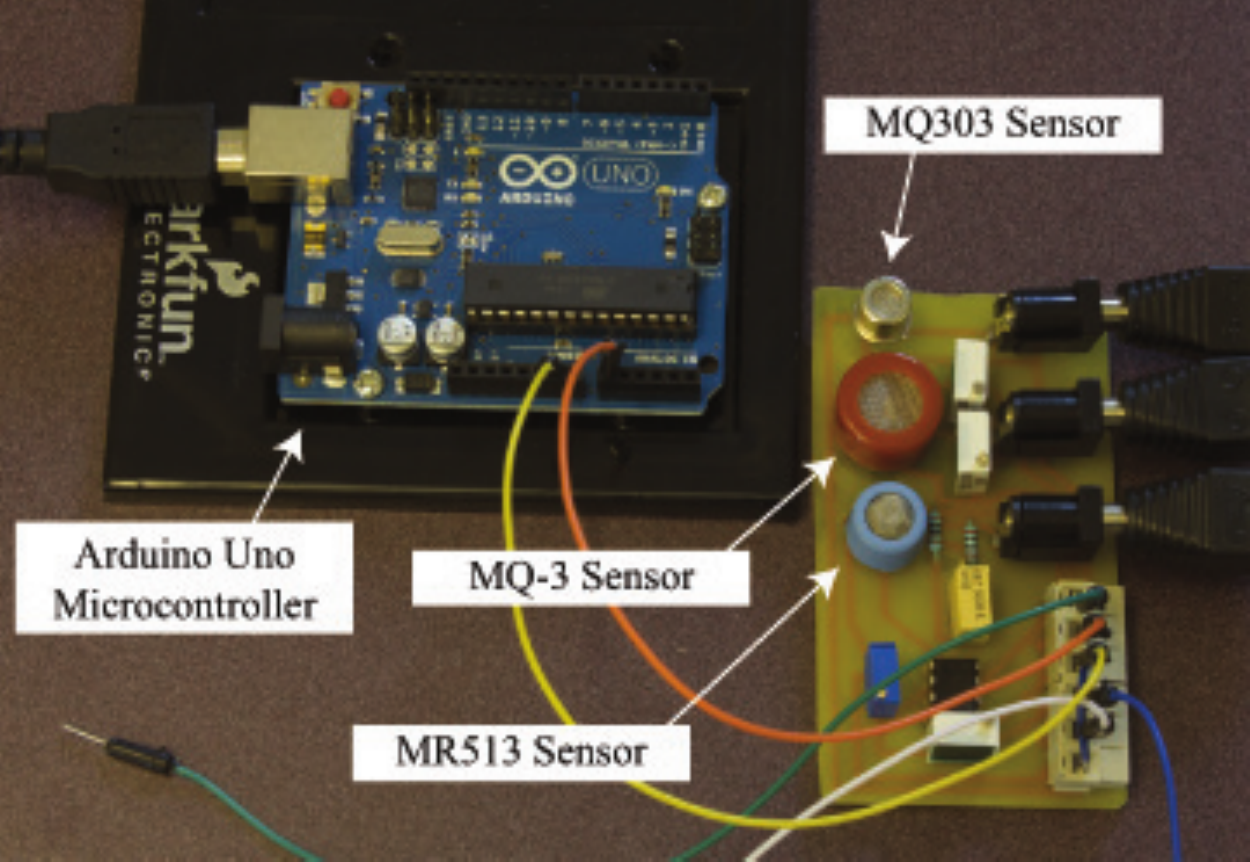}
	\end{center}
	\caption{\label{fig:RecPic} {\bf The receiver components.}}
\end{figure}

To achieve our design criteria, the receiver's sensor must be sensitive, widely available, and inexpensive. Moreover, it must be able to detect a volatile, widely available, and inexpensive signaling chemical that is safe at the low concentrations that we use. Therefore, we choose isopropyl alcohol (rubbing alcohol) as the signaling chemical with three different candidate sensors for demodulation and detection at the receiver: MQ-3, MQ303A, MR513 alcohol sensors, all of which are manufactured by Henan Hanwei Electronics Co. Ltd. of China. All three sensors use a metal oxide semiconductor detection layer for detecting the alcohol, but each has a different sensitivity, power and operation circuit diagrams. Besides isopropyl alcohol, the sensors can detect other types of alcohol such as ethanol. However, in this work we only use isopropyl. We implement all three sensors on a custom-made PCB board as shown in Figure \ref{fig:RecPic}.

\subsection*{Propagation Channel}
\label{sec:ChanProp}
We consider two different propagation schemes for the channel: diffusion, and flow assisted propagation. In the diffusion based propagation, after the initial spray the alcohol diffuses in the air until it reaches the receiver. In flow based propagation, a table-top fan is used to guide the alcohol towards the receiver. Therefore, the diffusion propagation does not require external energy (beyond the energy required to release the chemical message), while the flow assisted propagation requires external power. We use two different table-top fans to generate flow: 
\begin{itemize}
\item Honeywell 7 inch Personal Tech fan. This fan is an inexpensive balded fan (approximately \$16 USD) with two different fan speeds low and high.
\item Dyson AM01 10 inch bladeless fan. The Dyson fan is much more expensive (approximately \$250 USD), but can generate more laminar flows and many different wind speeds by adjusting an analog nub. 
\end{itemize}
When any of the two fans are used, they are placed 30 cm behind the spray.   

\begin{figure}[!h]
	\begin{center}
		\includegraphics[width=3.2in]{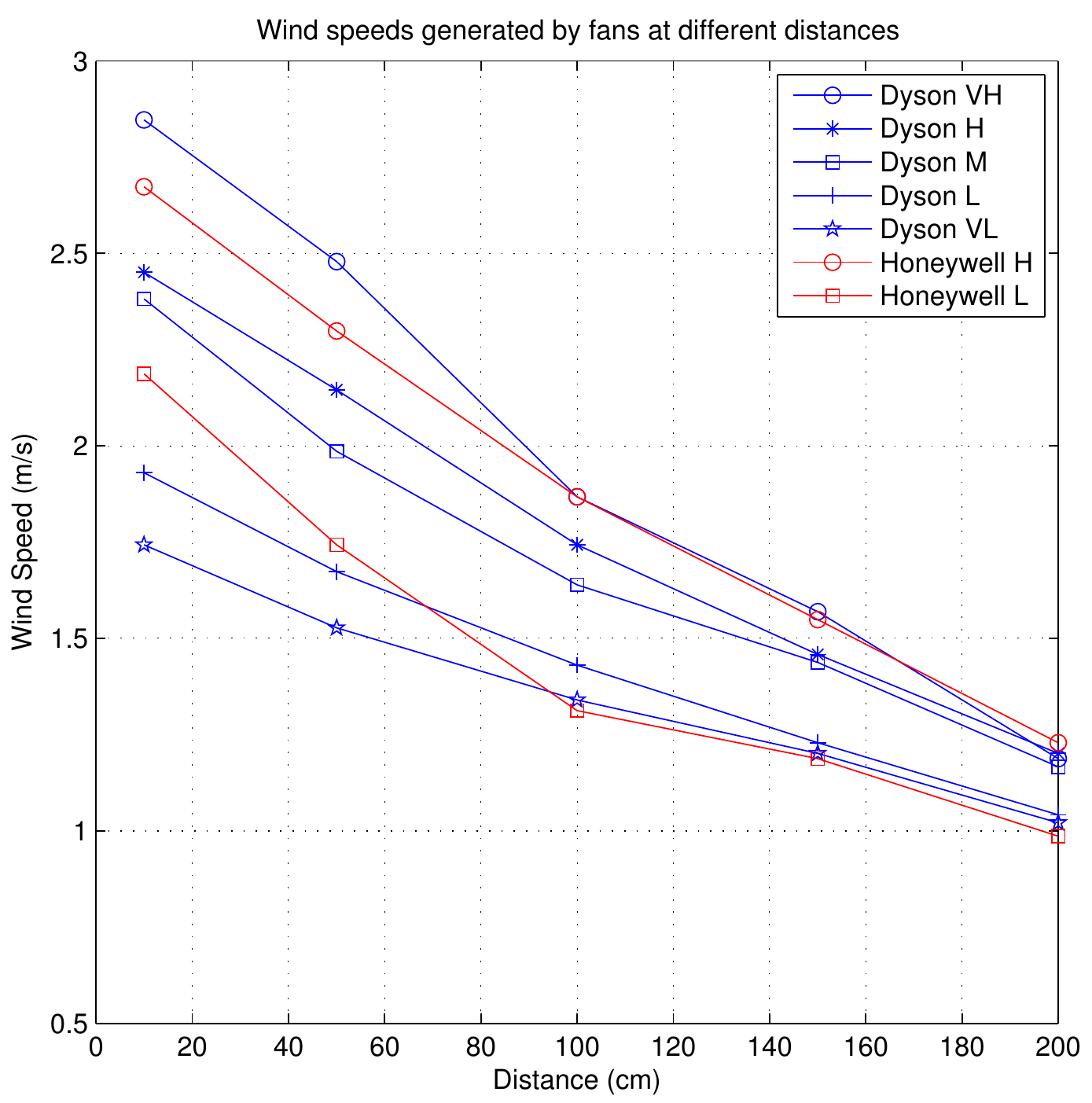}
	\end{center}
	\caption{\label{fig:fanSpeeds} {\bf Wind speeds generated by each fan.} The Dyson fan plots are in blue, and Honeywell plots are in red.}
\end{figure}

To measure the flow speeds generated by each fan we use the Pyle PMA82 digital anemometer. The maximum flow speed is measured at distances of 10 cm, 50 cm, 100 cm, 150 cm, and 200 cm from the front of the spray (the fan is placed 30 cm behind the spray). For the Dyson fan we select 5 different nub positions and we label these positions as very high, high, medium, low, and very low. Figure \ref{fig:fanSpeeds} shows the wind speed for each fan at each distance. Because there is +/$-$3\% error associated with our digital anemometer, we average four different measurements to produce each plot point in Figure \ref{fig:fanSpeeds}. Moreover, our digital anemometer is not rated for flow speeds below 1 m/s. Therefore, flow speeds below this range are not shown. The average flow velocities achieved over this distance is tabulated in Table \ref{tb:avgFlowVel} for each fan setting. 
\begin{table}[!h]
\caption{{\bf Average flow velocities.} The average flow velocities over the distance of 200 centimeters generated using the Dyson and Honeywell fans.}
\begin{tabular}{|l|c|}
\hline 
{\bf Flow Generated By} & {\bf Average Flow Speed (m/s)}  \\ 
\hline 
Dyson on very high & 1.99 \\
\hline
Dyson on high & 1.80 \\
\hline
Dyson on medium & 1.72 \\
\hline
Dyson on low & 1.46 \\
\hline
Dyson on very low & 1.37 \\
\hline
Honeywell on high & 1.92 \\
\hline
Honeywell on low & 1.48 \\
\hline
\end{tabular}
\label{tb:avgFlowVel}
\end{table}

\begin{figure}[!h]
	\begin{center}
		\includegraphics[width=3.2in]{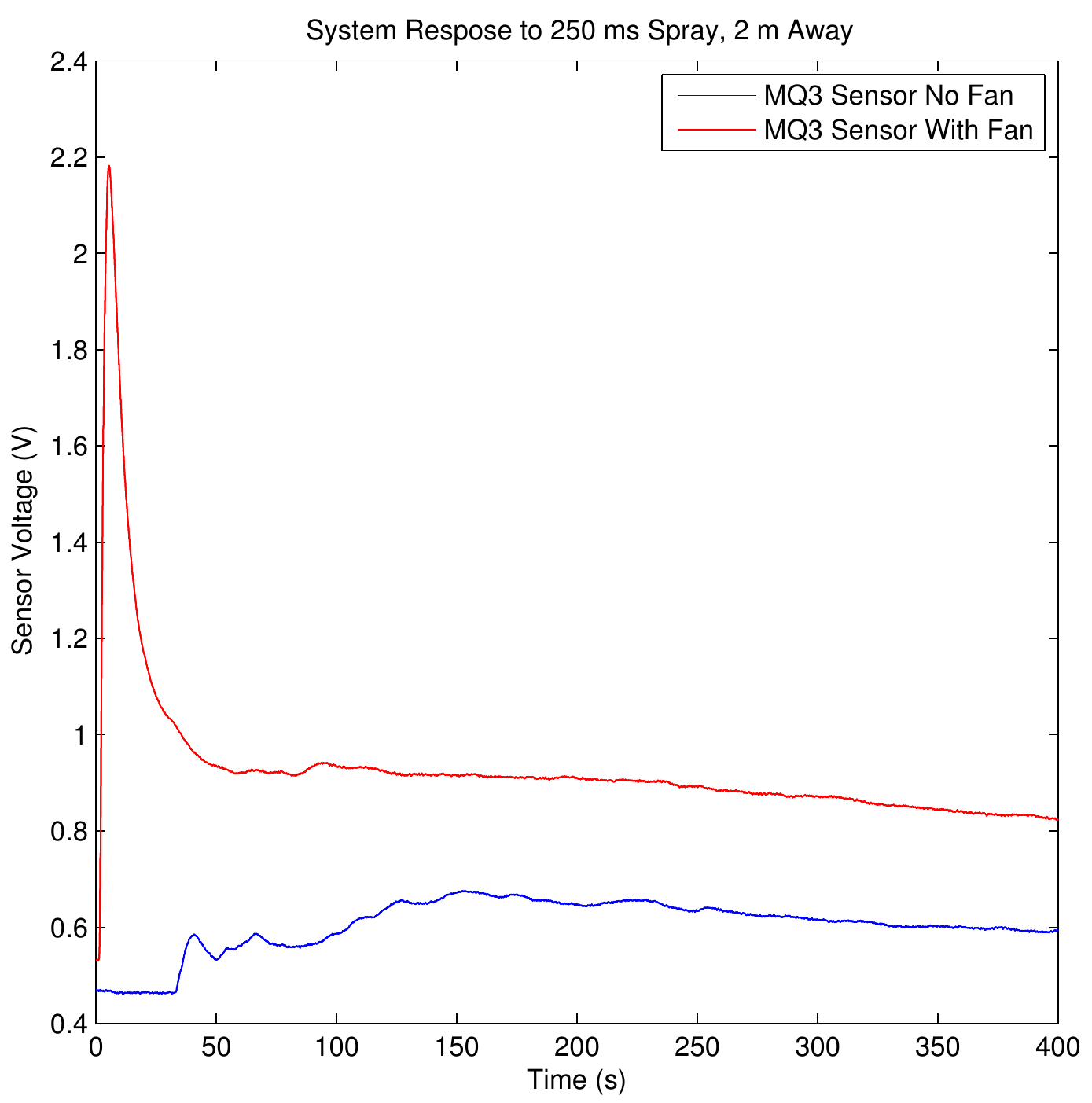}
	\end{center}
	\caption{\label{fig:fanVSnoFan} {\bf Diffusion based propagation versus flow based propagation.} The system response to a 250 ms spray 2 meters away for diffusion based propagation (blue), and flow based propagation (red).}
\end{figure}

We compared the system response (the output of the sensor for a single short spray) under both propagation schemes. At short distances (up to 1 meter), the diffusion based propagation scheme performs well because the alcohol ejected from the spray  reaches the sensors almost instantaneously. However, if the spray is placed further away, diffusion based propagation would not be practical because of the extremely slow system response. This effect can be seen in Figure \ref{fig:fanVSnoFan}, where the system response to a very short spray of 250 ms in duration is plotted for both diffusion based and flow based propagations. The flow in this figure is generated using our Honeywell fan on the high setting, and the spray is placed at a distance of 2 meters from the detection sensor. As can be seen, the system has a quick and distinct response when flow based propagation is employed. Although we plot the response for only one of the sensors (MQ-3 sensor), the same effect was observed for all the other sensors, as well as when the Dyson fan is used instead. Therefore, for our molecular communication setup we use flow based propagation. 

\subsection*{Signal Modulation and Demodulation}
Because communication is performed through chemical signals, and a limited amount of signaling chemical can be stored in a container at the receiver, the modulation and demodulation scheme selected should minimize the amount of chemical used. The source coding scheme for encoding text messages, presented in previous sections, uses the least amount of ones in the 5 bit sequence for characters that have a higher rate of occurrence in the English text. For example, letters  ``E'' and ``T'' both have a single one in their 5 bit sequences. Therefore, we modulate the bit 1 with a single spray and we modulate the bit 0 with no spray. This modulation scheme effectively minimizes the amount of chemical used for communicating English text.

At the receiver the demodulation is performed by measuring the rate of change in concentration. If during a single bit's communication session the voltage reading from one of the sensors is increasing (i.e. the concentration of the chemical signal is increasing), then the signal is demodulated as the bit 1. Similarly, if the voltage reading from one of the sensors is decreasing (i.e. the concentration of the chemical signal is decreasing) the signal is demodulated as the bit 0. More details regarding the detection and demodulation process is provided later in the paper.

\subsection*{Communication Protocol Design}
In this section, we discuss the communication protocol between the transmitter and the receiver, and its implementation. In designing the protocol we use the criteria that the protocol must be simple, asynchronous (i.e. no synchronization is required between the transmitter and the receiver), and should work independent of the separation distance between the transmitter and the receiver (i.e. it should not only work for a predefined fixed distance between the transmitter and the receiver). 

\begin{figure}[!h]
	\begin{center}
		\includegraphics[width=2in]{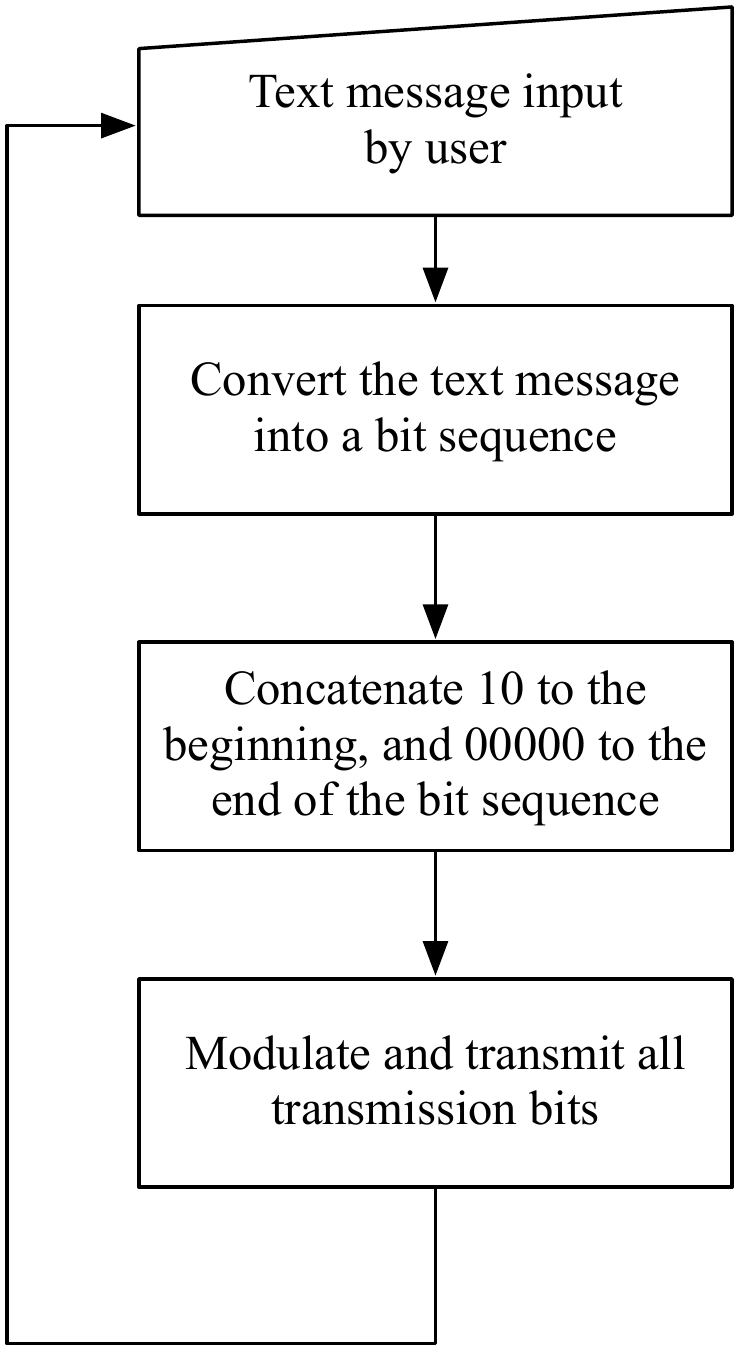}
	\end{center}
	\caption{\label{fig:TransFlowChart} {\bf Flow chart representation of the algorithm that controls the transmitter.}}
\end{figure}

At the transmitter, the output of the source encoder (i.e. the bit sequence representing the text message) is concatenated with a two bit sequence ``10'' at the beginning and the null character represented by ``00000'' at the end. The initial ``10'' indicates start of a text message and the null character indicates the end of the text message. For example, if the text message that is being transmitted is the letter ``A'', the output of the source encoder is the five bit sequence ``11000'' (where the left most bit position is the first bit position), and transmission bit sequence is ``101100000000''. The transmission bit sequence is then modulated using the scheme discussed in the previous section, where 1 is modulated with a spray and 0 with no spray. Figure \ref{fig:TransFlowChart} represents the flow chart of the algorithm that runs at the transmitter, and summarizes this process. 

At the receiver, there are two states: the {\em wait state}, and the {\em reception state}. In the wait state the receiver uses its sensor to continuously  monitor the concentration of alcohol. If there is a sudden increase in the concentration of alcohol (i.e. sudden increase in the sensor's voltage output), the receiver switches to the reception state. This sudden change is caused by the initial ``10'' bit sequence concatenated to the beginning of every text message sent by the transmitter. This sudden change can also be used as the reference time for synchronizing each bit interval for all the bits that would follow. Therefore, no synchronization is required between the transmitter and the receiver in advance. Another factor that is taken into account in this scheme is the propagation delay. Because the receiver is triggered into reception state as soon as the leading bit 1 is detected, the time delay caused by signal's propagation over the separation distance from the transmitter to the receiver is incorporated in the reference time. Therefore, the communication protocol is independent of the distance between the transmitter and the receiver, and it would work even when the distance is changed between communication sessions.

\begin{figure}[!h]
	\begin{center}
		\includegraphics[width=2in]{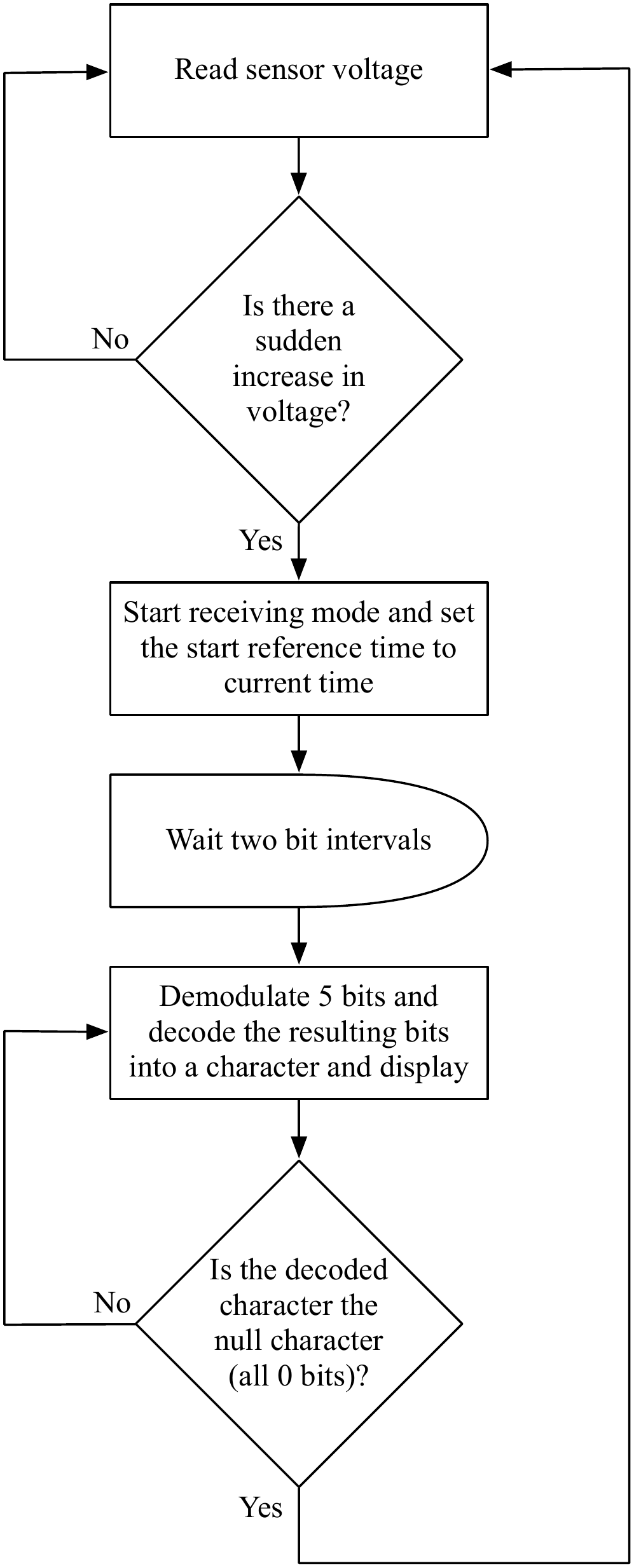}
	\end{center}
	\caption{\label{fig:RecFlowChart} {\bf Flow chart representation of the algorithm that controls the receiver.}}
\end{figure}
After the receiver enters the reception state, it waits for two bit intervals until the reception of the initial ``10'' bit sequence is finished. The receiver will then demodulates and decodes the received signal 5 bits at a time. During each 5 bit interval, the source decoded character is displayed to the computer screen using serial port connection. This process continues until the null character represented by all zero sequence ``00000'' is detected. Because the null character indicates the end of the text message, the receiver will go back to the wait state until another text message is sent by the transmitter. Figure \ref{fig:RecFlowChart} summarizes the algorithm that is implemented at the receiver.

\section*{Results and Discussions}
In this section we first present and discuss the impulse response of the overall system, and present the effects of different types of flow on the overall response. Based on these results, we choose the most suitable sensor to be used for our communication system. We then fine tune different system parameters such as transmission rate, and the demodulation/detection algorithm. Finally, some of the obtained results are presented and discussed. 

\subsection*{Overall System Response}
The overall system response is measured by using a very short spray that resembles the delta function from signal processing. Many parameters can effect the overall system response. The most notable factors that have a clear effect on the overall system response are:
\begin{itemize}
	\item The sensor: Each sensor has its own response to a changing concentration. We use three different sensors and choose the one that has the best overall response.
	\item The fan (flow type): Each fan has its own flow signature. We use both an inexpensive bladed fan and a bladeless fan to generate different types of flow at different flow velocities. The Dyson fan can produce a more laminar flow at various velocities. 
	\item The spray: Although we electronically control our spray with precise electrical signals, there are differences in the amount of particles that are released during each trial, and the size of the droplets in each spray stream. It is very difficult precisely control these parameter within an inexpensive apparatus. Therefore, we do not consider these effect on the overall system response and consider the overall response of the system to a very short spray.
\end{itemize}
There are other factors that could potentially effect the overall system response slightly such as other flow patterns within the room, room temperature, and humidity. To lessen the effects of these parameters, all the experiments are performed in a closed room with loosely regulated temperature and humidity. We believe these precautions are enough, because other factors have a much greater effect on the overall response.      

\subsubsection*{The Effects of the Sensor}

\begin{figure}[!h]
	\begin{center}
                \includegraphics[width=3in]{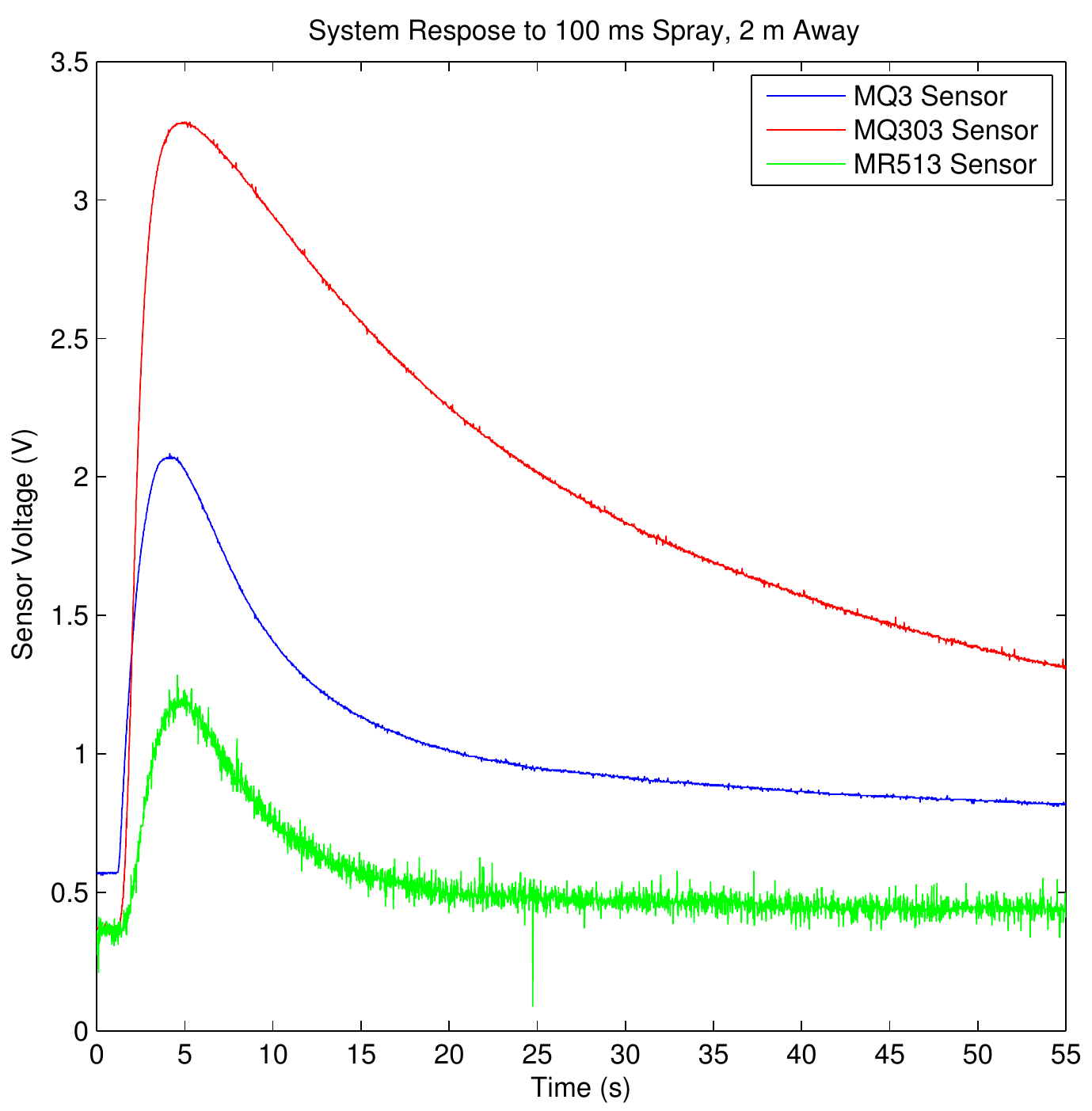}
    \end{center}
                \caption{\label{fig:DiffSenRes200} The system response to a short spray of 100 ms for all three sensors at 2 m separation between the transmitter and the receiver.}
                
\end{figure}

To study the effects of sensor on the overall response, the spray duration is set at 100 ms (i.e. the spray is switched on for 100 ms), and the system response is measured using each of the three sensors at various separation distances between the transmitter and the receiver. We use this scheme as it would be difficult to control and measure the actual volume of alcohol released during each burst. Figure \ref{fig:DiffSenRes200} and \ref{fig:DiffSenRes400} show the system response for 2 meter separation distance (\ref{fig:DiffSenRes200}), and 4 meter separation distance (\ref{fig:DiffSenRes400}) for all three sensors. The Honeywell fan on the high setting is used to produce the flows for all these plots. As expected the amplitude of the peak decreases and the delay before the peak increases as the separation distance is doubled. The peaks' full-width at half max also increases as the separation distance increases. Similar effects are also observed when the Dyson fan is used to generate the flow.  
\begin{figure}[!h]
\begin{center}
       \includegraphics[width=3in]{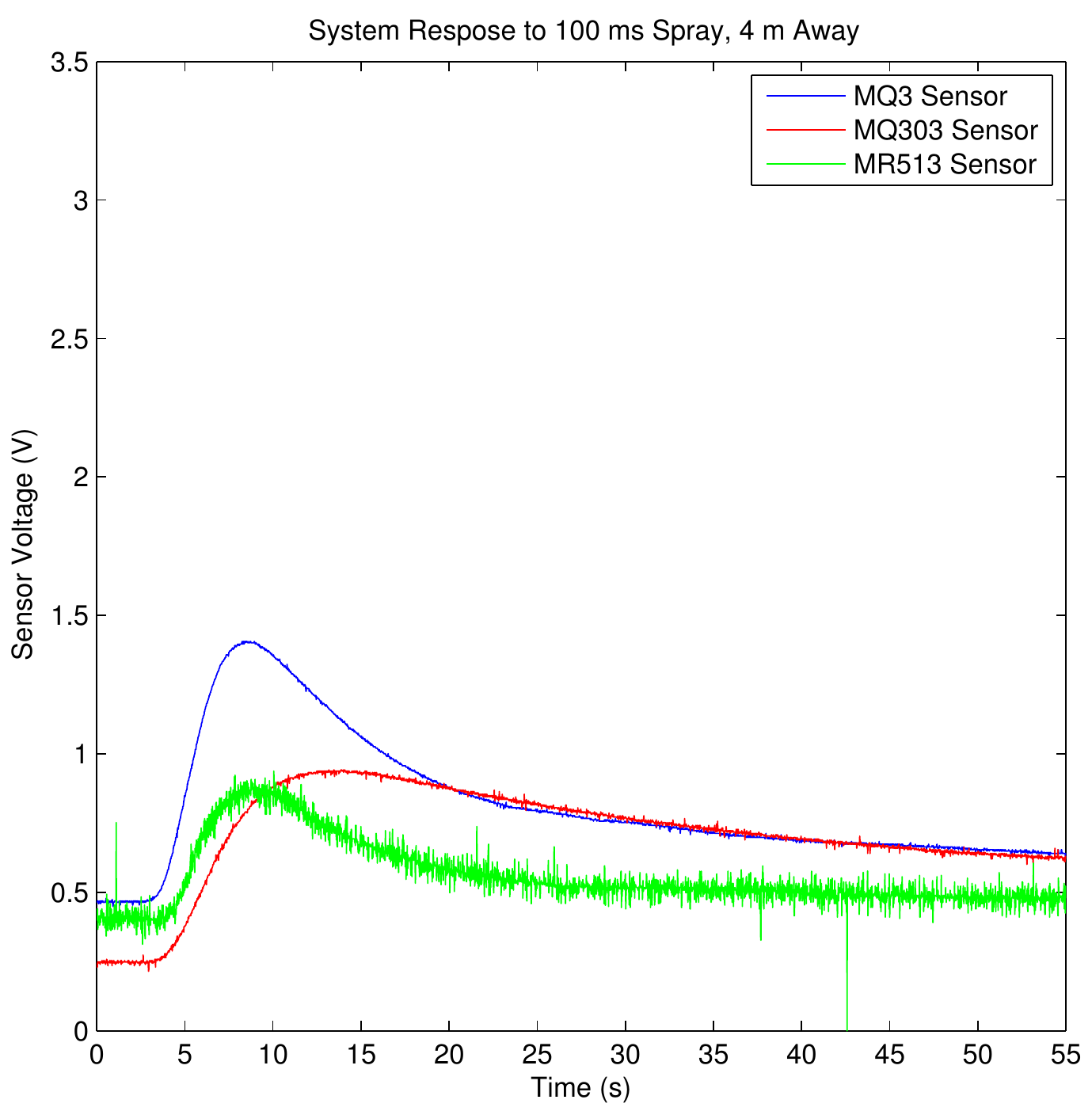}
       \end{center}
   \caption{\label{fig:DiffSenRes400s} The system response to a short spray of 100 ms for all three sensors at 4 m separation between the transmitter and the receiver.}
\end{figure}

From the overall system response, it is evident that there is a large amount of noise in the MR513's signal because of the operational amplifier used as part of its circuitry.  Although the MQ303 has a high peak at 2 meters, the peak's full width at half max is much larger than the other two sensors. Ideally, this width must be as small as possible. Moreover, the height of the MQ303's peak drops significantly at 4 meters. The MQ-3 sensor has low noise and better system response over wider range of separation distances. Moreover, the MQ-3 has the simplest circuitry and can draw power directly from the Arduino microcontroller board. Therefore, we select the MQ-3 sensor for our final implementation. 

\begin{figure}[!h]
	\begin{center}
		\includegraphics[width=3in]{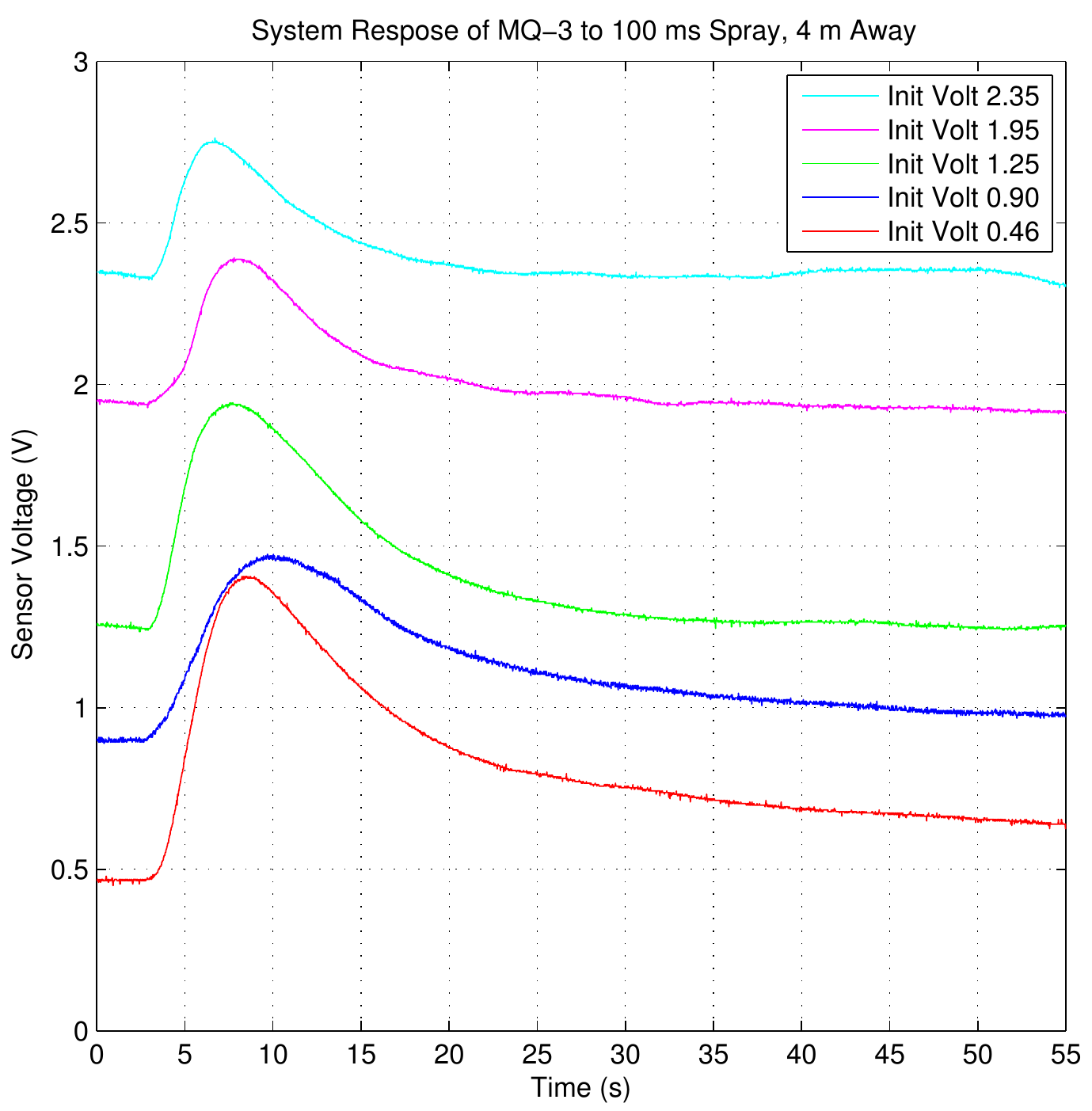}
	\end{center}
	\caption{\label{fig:DiffInitV400} {\bf System response for different initial voltages.} The system response changes based on the initial voltage.}
\end{figure}

One of the major caveats of metal oxide gas sensors, including all three sensors we consider in this paper, is the delays in responding to a changing concentration \cite{boc10}. These delays core categorized as: 
\begin{itemize}
	\item The sensor's {\em response time}, i.e., the time it takes for the sensor to respond to a change in concentration; and
	\item The sensor's {\em resume time}, i.e., the time it takes for the sensor to be used reliably again after a change in concentration. 
\end{itemize}
The change in system response based on the initial voltage reading (i.e., initial concentration at the sensor) is another factor affecting these sensors. This effect can be seen in Figure \ref{fig:DiffInitV400}, where the system response to a single short spray of 100 ms at the distance of 4 meters away with different initial voltage readings (i.e. different initial concentrations) at the sensor is plotted. The flow in this figure is generated using the Honeywell fan on high setting (similar results are also observed when the Dyson fan is used for flow generation).  This figure shows that the system response changes for different initial concentration levels at the sensor. To make sure that the sensor resume time is not effecting the readings, we bring up the sensor to a voltage level (i.e. concentration level) higher than the target initial voltage. We then wait long enough for the voltage reading to drop to the target initial voltage level, which is also long enough to abolish the effects of sensor resume time. We then initiate the impulse spray and measure the system response.

\subsubsection*{The Effects of Flow}

\begin{figure}[!h]
	\begin{center}
		\includegraphics[width=3in]{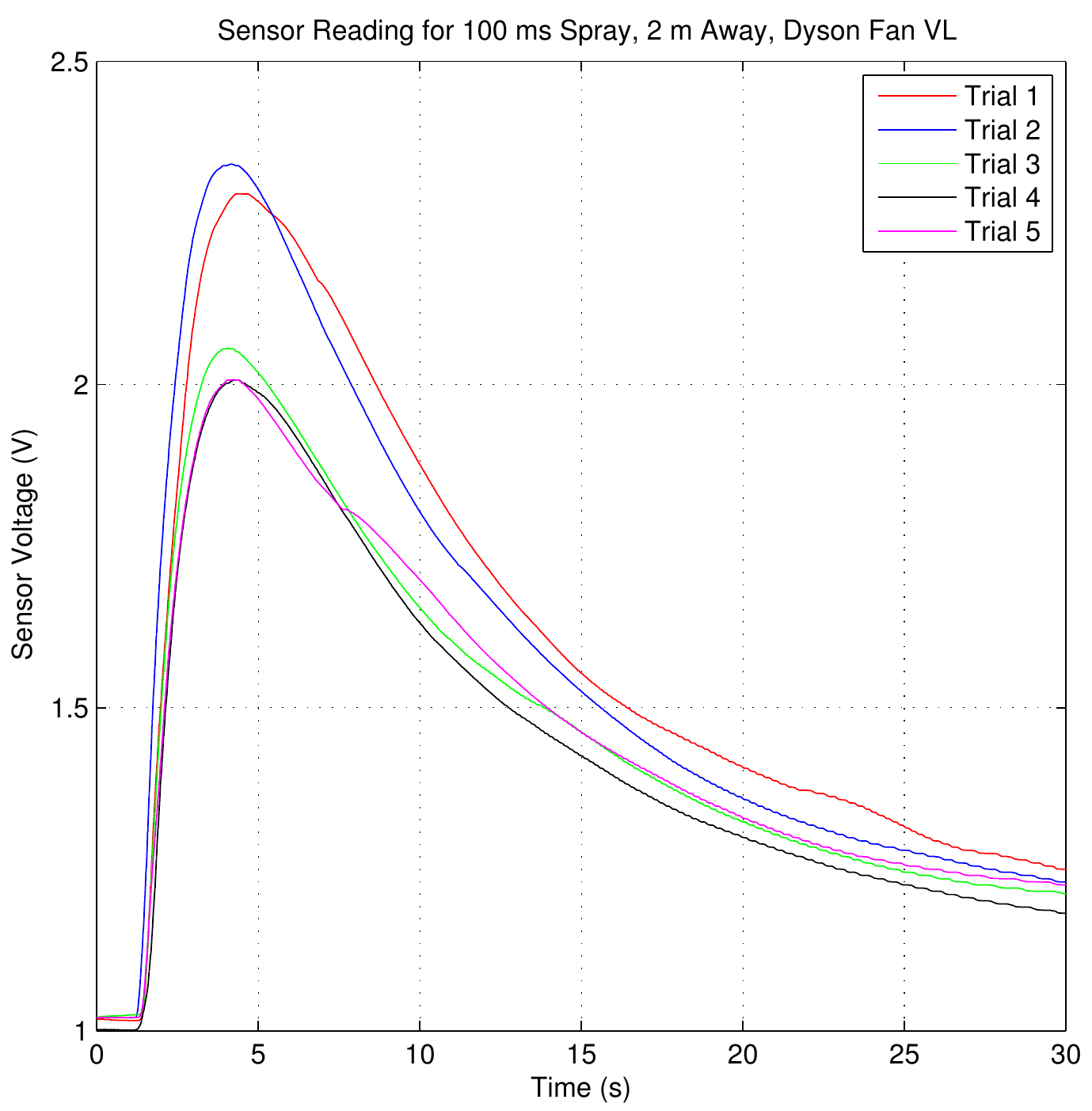}
	\end{center}
	\caption{\label{fig:DiffTrialsDVL} {\bf System response for different different experimental trials.} The flow in these trials is generated using the Dyson fan on very low setting.}
\end{figure}

Flow is an important part of our setup, because it carries the alcohol droplets from the transmitter to the receiver. Therefore, it has a significant effect on the overall system response. However, isolating the effects of flow can be very challenging. For example, the spray itself cannot release very precise amounts of alcohol with uniform droplet sizes between different experimental trials. Another factor that could potentially effect the results is random flows within the room. As a result, the overall impulse response of the system changes between different trials. This effect can be seen Figure \ref{fig:DiffTrialsDVL}, where the system response to a 100 ms spray 2 meters away is plotted for 5 different experimental trials. The initial voltage for each trial is kept constant at 1.02 volts, and the Dyson fan on the very low setting is used to generate the flow. From the plot it is evident that there is some difference across trials.

To mitigate this problem, and further isolate the effects of flow, we perform multiple trials and average the results. We use two performance measures for comparing the system response generated using different fans and flow speeds: the peak's max to full-width at half max (PMFWHM), and delay to peak's max (DPM). The PMFWHM is the ratio of the peak's maximum voltage to the full width of the peak at half max. The larger this ratio the taller and narrower the peak shape will be. Ideally the peak must be as tall and as narrow as possible. Therefore, larger ratios are desirable. The DPM, is the time from the start of the spray to the time where the peak's maximum point is achieved. The smaller this time duration, the faster the peak's maximum is reached. It is desirable for this delay be as small as possible. 

As explained earlier we use two different table fans made by Dyson and Honeywell. The Dyson fan is bladeless, more expensive, and can create more laminar flows. The Honeywell fan is inexpensive but it is a bladed fan and it creates more turbulent flows. Five different fan settings are considered for the Dyson fan, while the Honeywell fan has only two possible settings as explained in previous sections. The average flow velocities over a 2 meter distance are tabulated in Table \ref{tb:avgFlowVel}. For each fan and each corresponding fan setting, the overall system response to a short spray of 100 ms, 2 meters away, is measured for 10 experimental trials. The initial sensor voltage reading for each trial is kept constant at about 1.02 volts (i.e. there is enough delay between trials such that the sensor voltage falls back to 1.02 volts). 
\begin{figure}[!h]
   \begin{center}
   		\includegraphics[width=3in]{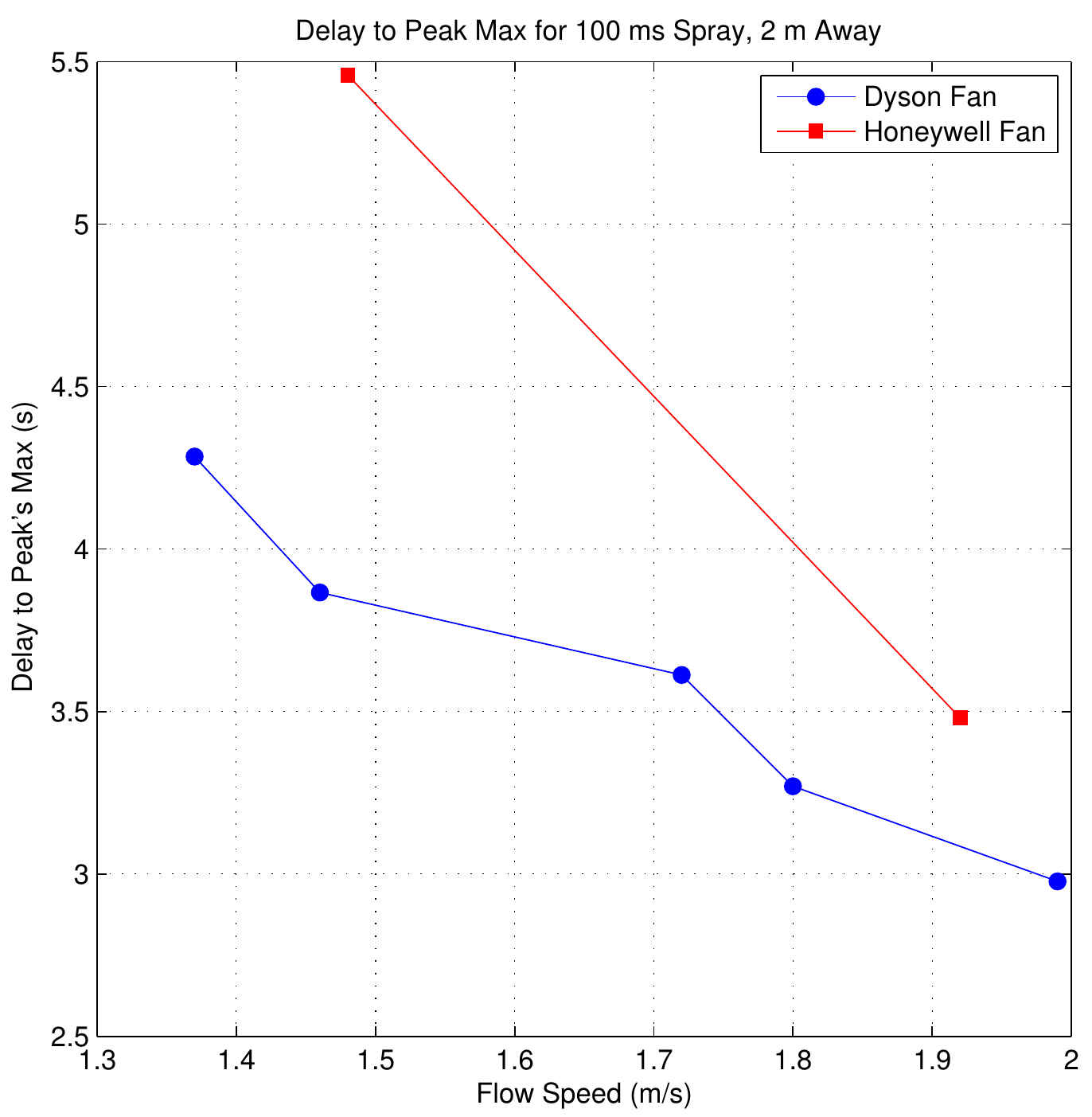}
   \end{center}
    \caption{\label{fig:flowDelayMax} The delay to peak's maximum measure for different flows.}  
\end{figure}
\begin{figure}[!h]
   \begin{center}
   	  \includegraphics[width=3in]{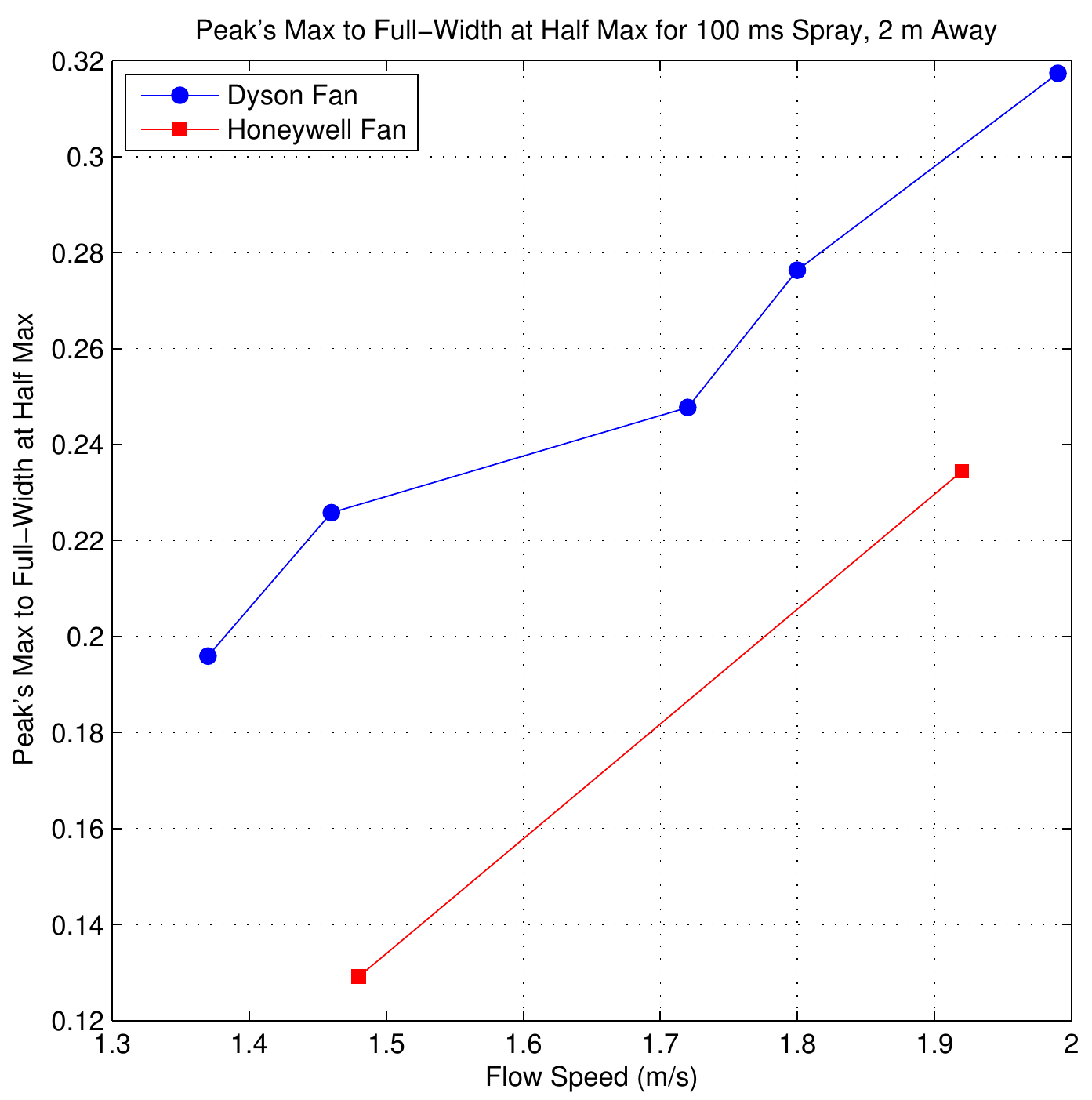}
   \end{center}
   \caption{\label{fig:flowPMFWHM} The peak's maximum to full width at half max measure for different flows.}
\end{figure}

To compare the fans and their corresponding fan settings, measures, PMFWHM and DPM, are calculated for each of the 10 trial. The results  are then averaged and presented in Figures \ref{fig:flowDelayMax} and \ref{fig:flowPMFWHM}. In Figure \ref{fig:flowDelayMax} the DPM is plotted for different flow speeds generated by each fan and its corresponding fan setting. As can be seen, the Dyson fan has a shorter delay to peak's max for the same average flow speed because the flow generated by this fan is more laminar compared to the Honeywell fan. Moreover, the flow speed also decreases the delay. Finally, from the Dyson plot we can see that this delay decreases almost linearly with increasing flow speed. 

The PMFWHM is shown in Figure \ref{fig:flowPMFWHM} for different fans and setting. From the plots we can see that the PMFWHM ratio increases as the fan speed increases. Therefore, the impulse response becomes narrower and taller as the fan speed increases. The Dyson fan also achieves higher ratios compared to the Honeywell fan. Therefore, the more laminar flows that the Dyson fan generates can create taller and narrower system response. 

From these results, we conclude that the Dyson fan is a better choice for generating flows. However, because it is more than 10 times expensive compared to the Honeywell fan, and one of our goals is to create a cost effective demonstration of macroscale molecular communication, for our final communication system we use the Honeywell fan. Therefore, in our final system the achievable transmission rates can be considered as a ``lower bound'', where it can be improved by simply using the Dyson fan.

\subsubsection*{System Nonlinearity}
 
\begin{figure}[!h]
	\begin{center}
		\includegraphics[width=3in]{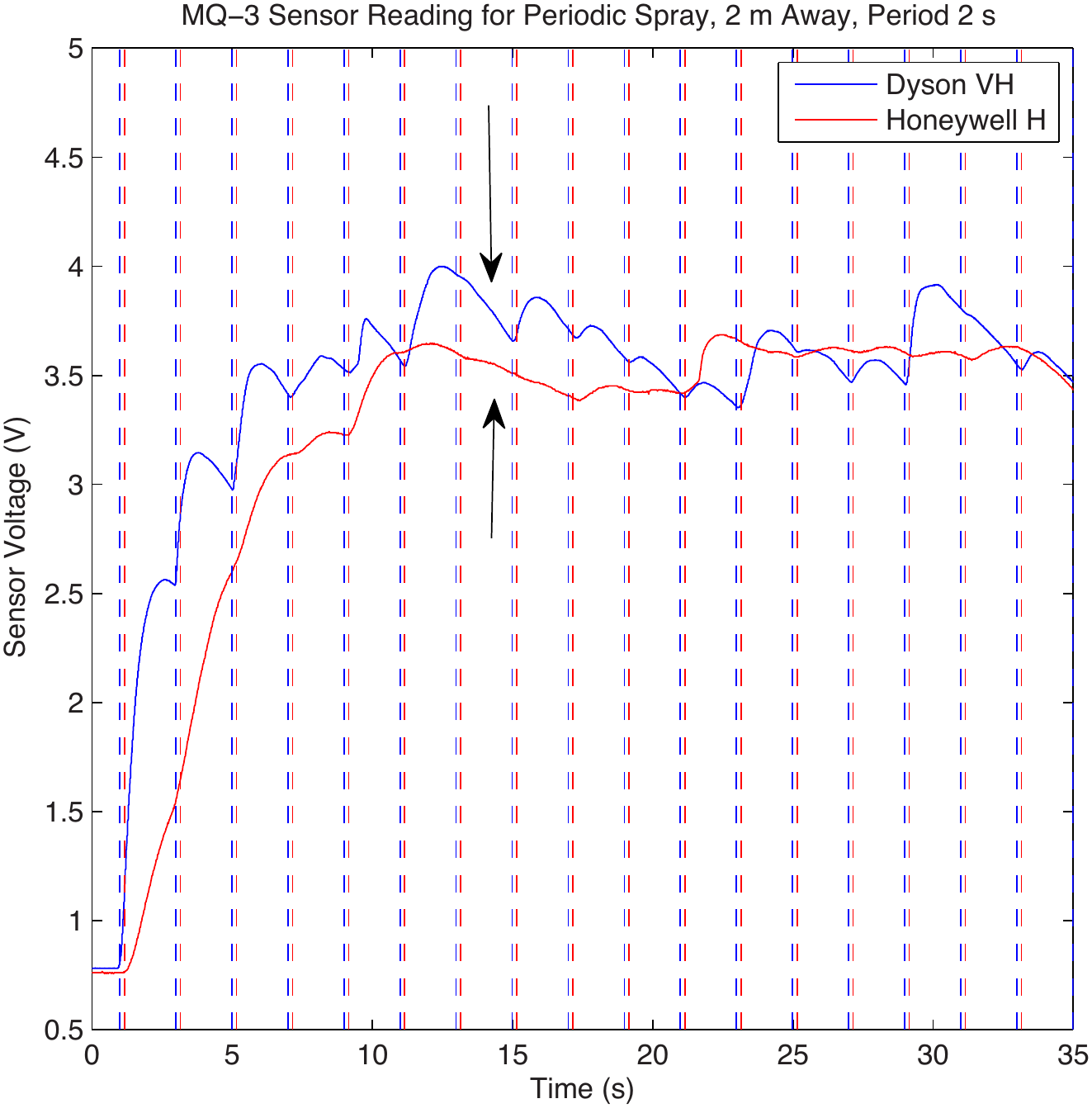}
	\end{center}
	\caption{\label{fig:PerSigRes200} {\bf System response to a periodic spray.} The dashed lines are 2 seconds apart and show each period. The arrows show the location where the sensor voltage decreases instead of increasing. }
\end{figure}

In this section we show that the overall system response of our setup is nonlinear. Although finding the exact cause of the nonlinearity is not possible, and more extensive research is required, this result by itself is very interesting. To show that the system is nonlinear, we consider a set of periodic sprays, of 100 ms with a period of 2 seconds. The output sensor voltage is them measured and recorded as the system response. Figure \ref{fig:PerSigRes200} shows the results for both the Dyson fan on the very high setting and the Honeywell fan on the high setting.    

As can be seen in Figure \ref{fig:PerSigRes200}, the output does not follow that of a linear system. For example, at about 13 and 15 seconds (the arrows in the plot point to this time duration), where there should be another increase in concentration because of the sixth and seventh periodic spray, there is a sudden drop in voltage. This effect is observed for both the the case where the Dyson fan is used and the case where the Honeywell fan is used. However, there are more clear peaks when the Dyson fan is used because of the narrower and taller system response explained in the previous section.

The nonlinear responses observed in our experiments are surprising, because most molecular communication systems are normally assumed to be linear in the literature. Many of the mathematical tools used in the literature at microscales require that the system is linear, and these tools cannot be directly applied to a nonlinear communication system. Although the source of nonlinearity is not known, some likely candidates are: the sensor with its response and resume times, the flow generated by our fans which may be turbulent, the spray which is not precise enough to create uniform streams, and other environmental factors such as other flows within the room. It may be possible that with more expensive and sensitive hardware, and within a precisely controlled environment the system response would be linear. Nonetheless, the potential nonlinearity of our system is an issue which merits further investigation.

\subsection*{Final Implementation and Discussion}

The final steps of implementation are discussed in this section. To make our system more cost effective, in our platform we use the Honeywell fan for demonstration despite the fact that the Dyson fan can create better system response. Although both fans can be used in our setup and slightly higher data rates can be achieved with the Dyson fan. We also use the MQ-3 sensor because it provides the best system response and it has the simplest circuitry between all three sensors. 

First, we address the issue of noise. Although the MQ-3 sensor response is less corrupted by noise compared to the other two sensors, there is still some noise present in the signal. To further reduce this noise, 20 ms of sensor data is averaged to generate a single sensor reading. Because the Arduino sampling rate is observed to be about 8.33 kHz, 20 ms of sensor data contains 167 different readings which are then averaged. Therefore, in the wait state the receiver checks consecutive 20 ms of averaged sensor readings, and triggers a change to the reception state if the difference between the current reading and the previous reading is greater than 0.5 levels (because Arduino has a 10 bit analog to digital converter the sensor reading would be an integer between 0 and 1023 representing 1024 different voltage levels, where 0 represents 0 volts and 1023 represents 5 volts).

An important communication parameter is the transmission rate. One of the major factors that affects reliable communication at a given transmission rate is the DPM. The DPM is in turn affected by the flow type and the flow speed. Therefore, for our platform the fan speed is always set to high. Another factor that effect the transmission rate is the sensor response and resume times discussed in the previous section. Finally, many other factors such as the environmental noise (e.g. random flow patterns in the room) can also effect the transmission rate.
\begin{table}[!h]
\caption{{\bf Different transmission rates and their reliability.}}
\begin{tabular}{|l|c|c|c|}
\hline 
 & \multicolumn{3}{ c| }{{\bf Distance}} \\
 \hline
{\bf Transmission Rate (bits/s)} & {\bf 2 m} & {\bf 3 m} & {\bf 4 m} \\ 
\hline 
0.2 & Very Reliable & Very Reliable & Very Reliable \\
\hline
0.33 & Reliable & Very Reliable & Reliable \\
\hline
0.5 & Unreliable & Unreliable & Unreliable \\
\hline
\end{tabular}
\label{tb:TransReliab}
\end{table}

We tried various transmission rates from one bit per 5 seconds (a character per 25 seconds) to one bit per 2 seconds (a character per 10 seconds). To measure the reliability at each rate, we performed multiple experiments at different separation distances between the transmitter and the receiver. We then classified each transmission rate at each separation distance according to the following ranking: very reliable (bit error rates of less than 0.01), reliable (bit error rates of 0.01 to 0.03), unreliable (bit error rates greater than 0.03). Table \ref{tb:TransReliab} summarizes the results.

At one bit per 2 seconds the transmission is unreliable at small distances of up to 2 meters, because of the sensor's resume time at higher concentrations is longer. Moreover, at larger distances (greater than 2 meters) successful communication is not possible at the rate of one bit per 2 seconds. At the rate of one bit per 5 seconds, the transmission is very reliable over various separation distances from 4 meters to 1 meters. Based on experiments, the smallest transmission rate that is reliable at distances up to 4 meters is one bit per 3 seconds. At this rate over the separation distance of 4 meters the communication session is reliable, over the separation distance of 3 meters the communication session is very reliable, and over the separation distance of 2 meters the communication session is reliable. The reason that the communication channel degrades slightly as the separation distance is decreased from 3 meters to 2 meters is because of the higher concentration levels at the sensor, and hence longer sensor resume times.

\begin{figure}[!h]
	\begin{center}
		\includegraphics[width=3in]{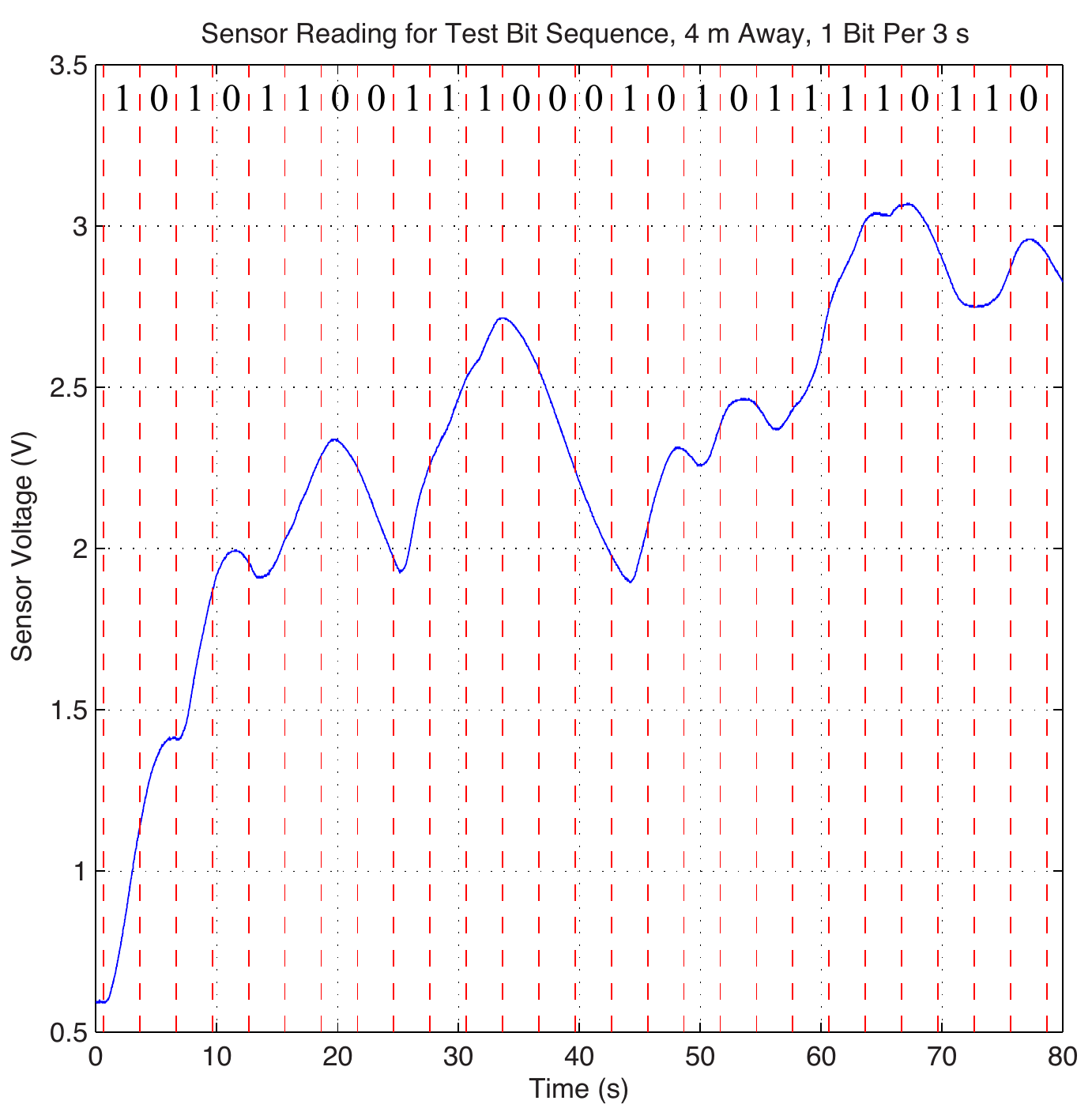}
	\end{center}
	\caption{\label{fig:TestSeq} {\bf Received signal when a 26 bit test sequence is transmitted.} Sensor reading for the 26 bit test sequence ``10101100111000101011110110'' transmitted 4 meters away at the rate of one bit per 3 seconds. The dashed red lines represent the start and the end of each bit. }
\end{figure}
In the rest of this section we focus on this transmission rate (one bit per 3 seconds), and describe in details the demodulation and detection algorithm for this rate. Although this algorithm is slightly different for each transmission rate, the same underlying principal is used for detection and demodulation at all rates: the rate of change in the concentration at the sensor. To fine tune this algorithm a 26 bit test sequence ``10101100111000101011110110'' is transmitted at the distance of 4 meters away, and the sensor reading is recorded. Figure \ref{fig:TestSeq} plots the sensor voltage reading during this transmission session. Dashed red lines are used to represent the start and the end of each bit. From this plot we devise a simple detection and demodulation scheme. The difference between the voltage level (there are 1024 levels in the Arduino's 10 bit analog to digital converter) at the end of a bit interval and the middle of a bit interval is measured. If the difference is greater than 2.2 levels (this threshold is derived through experimentation), the bit is detected as 1; otherwise the bit is detected as 0.    

\begin{figure}[!h]
   \begin{center}
   		\includegraphics[width=3in]{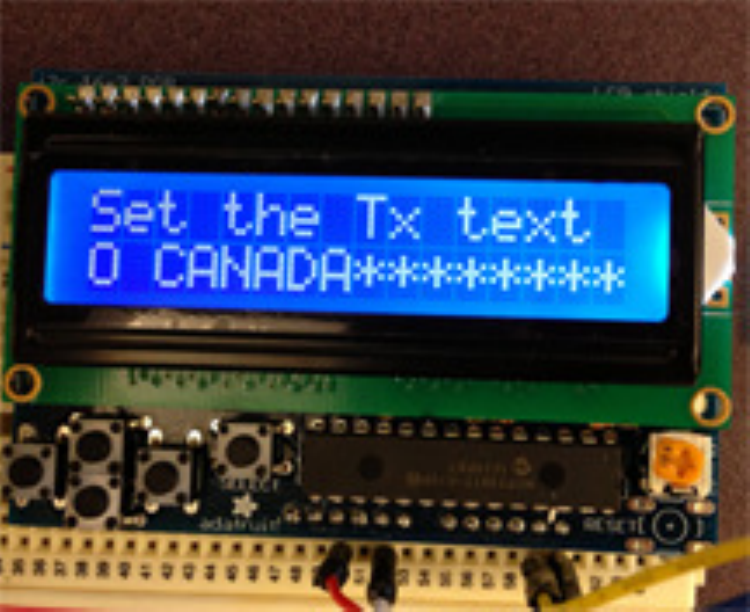}
   \end{center}
   \caption{\label{fig:TransTxt} The text entered at the transmitter.} 
\end{figure}

\begin{figure}[!h]  
   \begin{center}
   		\includegraphics[width=3in]{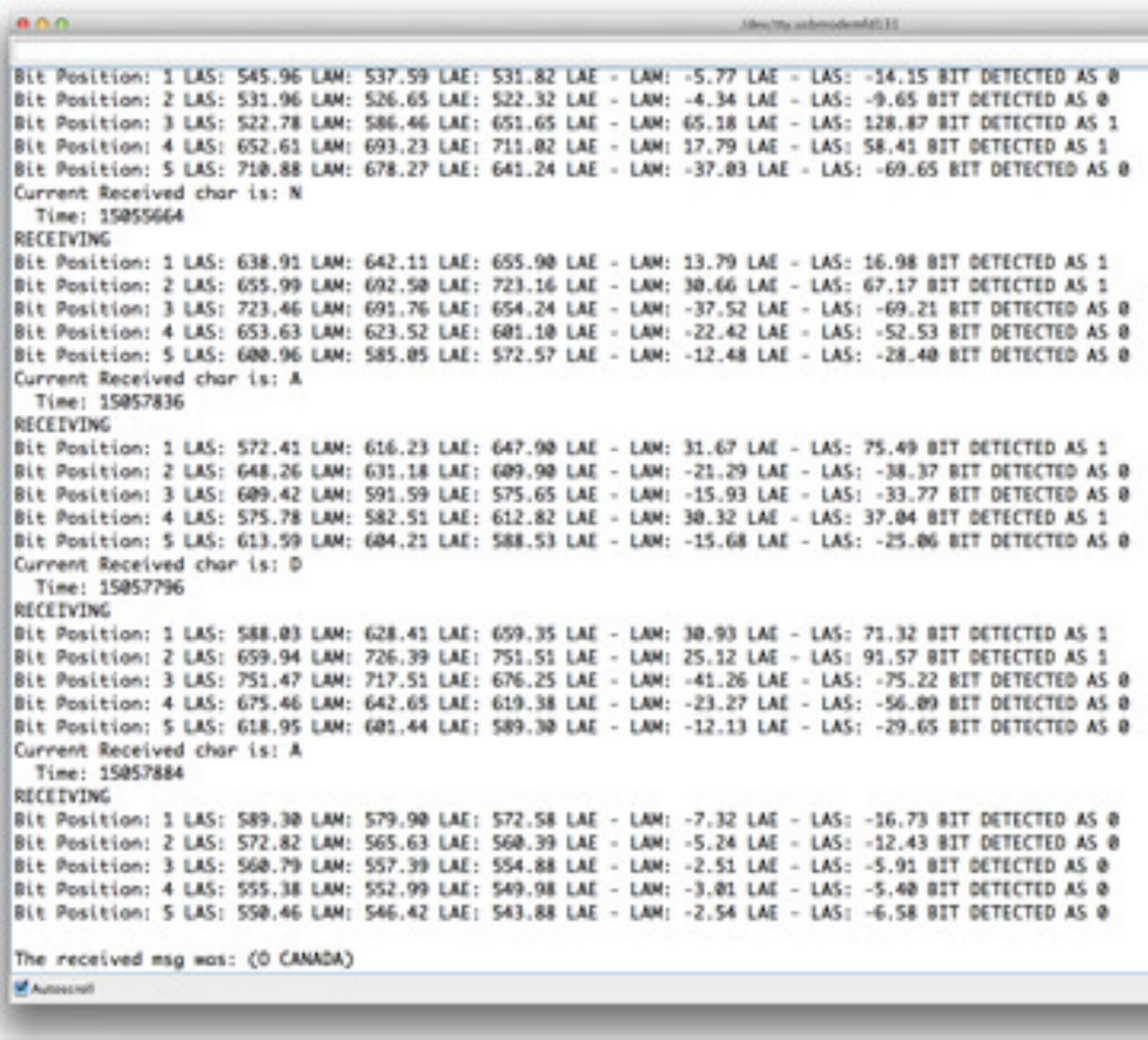}
   \end{center}
   \caption{\label{fig:RecTxt} The text received at the receiver.}
\end{figure}

Using this scheme, we are able to successfully transfer the test phrase ``O CANADA'' (the name of the national anthem of Canada) from the transmitter to the receiver. Figures \ref{fig:TransTxt} and \ref{fig:RecTxt} show this test phrase at the transmitter and received at the receiver. There are a number of multimedia files accompanying this paper that show the text entry process at the transmitter, and multiple communication sessions at this transmission rate over different separation distances.

\section*{Conclusions and Future Work}

In this work we developed the first known platform capable of transmitting short text messages using chemical signals. Our goal was to keep the communication protocol and algorithms simple such that other researchers could replicate these results. Moreover, the sensors and equipment were chosen to be inexpensive and widely available. Therefore, our first major contribution was the development of the platform itself, which demonstrated the feasibility of molecular communication at macroscales. Another purpose of this platform was to motivate future research and bridge the gap between theory and practice. 

After carefully selecting the necessary materials for our platform, we analysed the overall system response of our setup. We showed that there is a linear relationship between flow speed and the delay to system response peak's maximum. We showed that there is also a linear relationship between the flow speed and the peak's maximum to full-width at half max. Moreover, we demonstrated that more laminar flows have better narrower system response, which is desirable. 

Another major finding was the nonlinearity of our platform. This finding is very important because most of current communication theory is based on linear system. Although we were unable to find the exact reason for nonlinearity, we provided some guide posts. This motivates further study on the exact cause of the nonlinearity in future works. If it is shown that the nonlinearity is part of practical molecular communication systems (i.e. the nonlinearity cannot be resolved using better equipment), new communication theoretic work may be necessary on this topic.   

In the final part of the paper, we demonstrated a practical molecular communication system capable of transmitting short text messages. Although high transmission rates were not achieved in this work, the transmission rates can be significantly improved by using better fans, more sophisticated protocols and detection algorithms, use of multiple chemicals, use of multiple-input and multiple-output (MIMO) communication, designing better sensors, and using sensor arrays. We leave these for future study.


\bibliography{paper}




\end{document}